\documentclass{elsart4-1}

\usepackage{amssymb}

\usepackage[english,francais]{babel}

\newtheorem{theorem}{Theorem}[section]

\newtheorem{e-proposition}[theorem]{Proposition}

\newtheorem{e-definition}[theorem]{Definition\rm}
\newtheorem{remark}{\it Remark\/}

\renewcommand{\theequation}{\arabic{equation}}
\def\og{\leavevmode\raise.3ex\hbox{$\scriptscriptstyle\langle\!\langle$~}}
\def\fg{\leavevmode\raise.3ex\hbox{~$\!\scriptscriptstyle\,\rangle\!\rangle$}}

\newcommand\otimesal{\mathop{\hbox{\raise 1.6 ex
  \hbox{$\scriptscriptstyle\mathrm{al}$}
\kern -0.92 em \hbox{$\otimes$}}}}
\newcommand\oplusal{\mathop{\hbox{\raise 1.6 ex
  \hbox{$\scriptscriptstyle\mathrm{al}$}
\kern -0.92 em \hbox{$\oplus$}}}}
\newcommand\Gammal{\hbox{\raise 1.7 ex
\hbox{$\scriptscriptstyle\mathrm{al}$}\kern -0.50 em $\Gamma$}}
\renewcommand\i{\mathrm{i}}

 \let\be=\beta \let\de=\delta 

  \let\ga=\gamma 
\let\ka=\kappa \let\la=\lambda  
\let\si=\sigma

\newcommand{\caB}{{\mathcal B}}
\newcommand{\caC}{{\mathcal C}}

\newcommand{\caF}{{\mathcal F}}

\newcommand{\caH}{{\mathcal H}}

\newcommand{\bbC}{{\mathbb C}}

\newcommand{\bbP}{{\mathbb P}}

\newcommand{\bbR}{{\mathbb R}}

\newcommand{\opunit}{\text{1}\kern-0.22em\text{l}}

\newcommand{\frh}{{\mathfrak h}}

\renewcommand{\d}{{\mathrm d}}
\newcommand{\sys}{{\mathrm S}}
\newcommand{\res}{{\mathrm R}}

\renewcommand{\sp}{\mathrm{sp}}

\newcommand{\beq}{ \begin{equation} }
\newcommand{\eeq}{ \end{equation} }
\newcommand{\bet}{ \begin{theorem} }
\newcommand{\eet}{ \end{theorem} }

\newcommand{\trace}{\mathrm{Tr}}

 \newcounter{smallarabics}
\newenvironment{arabicenumerate}
{\begin{list}{{\normalfont\textrm{\arabic{smallarabics})}}}
  {\usecounter{smallarabics}\setlength{\itemindent}{0cm}
  \setlength{\leftmargin}{5ex}\setlength{\labelwidth}{4ex}
  \setlength{\topsep}{0.75\parsep}\setlength{\partopsep}{0ex}
   \setlength{\itemsep}{0ex}}}
{\end{list}}

\newcounter{smallroman}

\newcommand{\ben}{\begin{arabicenumerate}}
\newcommand{\een}{\end{arabicenumerate}}

\begin{document}

\begin{frontmatter}


\selectlanguage{english}
\title{Quantum fluctuation theorem: Can we go from micro to meso?}


\selectlanguage{english}
\author[wojtek]{Wojciech De Roeck  }
\ead{wojciech.deroeck@fys.kuleuven.be}
\thanks[wojtek]{Postdoctoral Fellow of the FWO-Flanders}
\address{Instituut Theoretische Fysica, K.U.Leuven }
\address{Deptartment of Mathematics, Harvard University }



\begin{abstract}
Quantum extensions of the Gallavotti-Cohen fluctuation theorem
(FT) for the entropy production have been discussed by several
authors. There is a practical gap between  microscopic forms of FT
and mesoscopic (i.e.\ not purely Hamiltonian) forms for open
systems. In a microscopic setup, it is easy to state and to prove
FT. In a mesoscopic setup, it is difficult to identify
fluctuations of the entropy production. (This difficulty is absent
in the classical case.) We discuss a particular mesoscopic model:
a Lindblad master equation, in which we state FT and, more
importantly, connect it rigorously with the underlying microscopic
FT. We also remark that FT is satisfied by the Lesovik-Levitov
formula for statistics of charge transport.


\vskip 0.5\baselineskip

\end{abstract}
\end{frontmatter}

\selectlanguage{english}

\section{Introduction} \label{sec: introduction}
A goal of nonequilibrium statistical mechanics is to make good,
yet simple, models of complicated phenomena out of equilibrium and
to analyze them. Such models should answer correctly questions
like: ``What is the structure of an object's state that is kept
between two heat reservoirs at different temperatures? How does it
look like locally and how does it fluctuate?" or ``When and how
does a liquid flow become turbulent?". In equilibrium, we have a
good and simple guiding rule for making dynamical models:
\emph{detailed balance}. Surely, any nonequilibrium dynamics
should break detailed balance, but can one formulate a general
rule as to how to fix this detailed balance breaking?

It is here that the celebrated Gallavotti-Cohen fluctuation
theorem (FT) (see
\cite{evanscohen,gallavotticohen95prl,gallavotticohentwo}) for the
entropy production comes into the picture. It turns out that there
is such a general rule and it reads (for every path $\si$ of the
dynamics, with corresponding probability $\mathrm{Prob}[ \si]$)
\beq \label{rule} \log{ \frac{\mathrm{Prob}[ \si] }{\mathrm{Prob }
[\theta \si] }} = \textrm{Entropy production along } \si  + o(t)
\eeq where $\theta \si$ is the time reversed path $\si$. The FT
itself
 \beq \label{FT}
\log{ \frac{\mathrm{Prob }\left[ \textrm{Entropy production up to
time $t$ equals $s t$} \right] }{ \mathrm{Prob }\left[
\textrm{Entropy production up to time $t$ equals $-st$ } \right]}}
= s t + o( t) \eeq is an immediate consequence of the above rule
(\ref{rule}). The LHS are expected to be linear in time $t$, the
$o(t)$ term vanishes after dividing by $t$ and taking $t \uparrow
\infty$. See \cite{maesoriginanduse} for a discussion focusing on
the universality of (\ref{rule}).

For our purposes, it is good to  emphasize that there are two
reasons to appreciate  (\ref{rule},\ref{FT}), even if those
reasons are obviously intimately related:

  \begin{enumerate}


 \renewcommand{\theenumi}{\Roman{enumi}}
\item \label{item 1}Formulas (\ref{rule}) and (\ref{FT}) can be derived on a microscopic level, starting with
Hamiltonian dynamics, using its time-reversibility and identifying
entropy production as the $log$ of phase space volume. This was
outlined in \cite{maesnetocny}.
\item \label{item 2} The rules (\ref{rule}) and (\ref{FT}) do  apply to a lot of effective (not purely Hamiltonian) or mesoscopic
models.
 It is actually in effective models that
the FT was observed and that it is studied most often. In
\cite{evanscohen,gallavotticohen95prl,gallavotticohentwo}, one
treats thermostatted systems, while in e.g.\
\cite{kurchanstochastic,lebowitzspohn2,crooks}, as well as in most
later publications, one deals with stochastic models.
\end{enumerate}

Note that (\ref{item 2}) does morally, but not exactly follow from
(\ref{item 1}) since mesoscopic or effective models  are
\emph{merely} good approximations of the Hamiltonian microscopic
dynamics.

This article is concerned with the question how rules (\ref{rule})
and (\ref{FT}) extend to quantum mechanics,  On the microscopic
level (cfr.\ (\ref{item 1}) above), there is an obvious problem
with deriving (\ref{rule}); quantum mechanics does not have the
concept of paths with associated pathwise entropy production.
However, (\ref{FT}) still makes sense and in fact it remains true,
as was pointed out in
\cite{jarzynskiquantum,kurchanquantum,tasakijarzynski,deroeckmaes}
and as we review in a streamlined version in Section \ref{sec:
qft}.

As to the status of quantum FT in effective or mesoscopic models
(cfr.\ (\ref{item 2})), the majority of research has focused on
master equations for small systems (see recently
\cite{espositoharbolamukamel}). A very explicit discussion can be
found in \cite{deroeckmaesfluct}, a simplified version of which is
presented in Section \ref{sec: guessing}. The ideas developed
there are also implicit in \cite{deroeckmaes} and \cite{mukamel}.

The fact that not more effective or mesoscopic models have been
investigated is perhaps not too surprising. With the possible
exception of the master equations  discussed in Section \ref{sec:
master}, there are not much  Markovian models of open quantum
systems which are healthy in all respects. One frequent pathology
is the appearance of nonpositive density matrices. A review of
this problem and a list of possible approaches is given in
\cite{kohenmarston}. We should mention that  these problems  are
not inherent to nonequilibrium models. Already the notion of
detailed balance is somehow problematic for Markovian quantum
models since the existing definitions \cite{agarwal,alicki} are so
restrictive that the master equation in Section \ref{sec: master}
is \emph{essentially} the sole example.

 Moreover,
even if one has a trustworthy model, there still remains  to
identify the fluctuating entropy production. It should be clear
from Section \ref{sec: master} and Remark \ref{rem: unravel} in
Section \ref{sec: guessing} that this is not trivial.

We remark hence that  in quantum mechanics, there is a practical
gap between (\ref{item 1}) and (\ref{item 2}). This might be
disconcerting, but possibly it also offers opportunities: perhaps
one can improve upon some models of open quantum systems \emph{by
imposing FT}.

In the case of master equations which we treat in this article,
not only (\ref{FT}) holds but there is also a convenient framework
to state (\ref{rule}). This framework is the formalism of
\emph{unraveling of master equations}. Our results imply that
these unravelings are rigorously linked to fluctuations of
currents. This is discussed extensively in
\cite{derezinskideroeckmaes}.

\subsection{Brief summary}

In Section \ref{sec: qft}, we review and sharpen the Hamiltonian
approach to the FT. In Section \ref{sec: ll}, we give an example
of a known counting formula in mesoscopic physics which satisfies
FT. We introduce the master equation describing a small system in
contact with different heat baths in Section \ref{sec: master} and
we develop the necessary tools (unravelings) to identify the
entropy production. Finally, in Section \ref{sec: result}, we
connect the master equation with the Hamiltonian approach and we
show that fluctuations of the entropy production in the
Hamiltonian model converge to fluctuations calculated by means of
unravelings. The importance of this result is: \ben
\item { It shows that the FT is valid in the thermodynamic limit
(at least in the weak coupling limit)}.
\item{  It rigorously justifies the technique of unravelings, which
is quite popular in quantum optics.} \een

\section{The Quantum fluctuation Theorem  } \label{sec: qft}

To fix thoughts, we introduce a model and corresponding notation.
Imagine several heat reservoirs  indexed by $k \in K$ and modeled
by Hilbert spaces $\caH_k$ and self-adjoint Hamiltonians $H_k$ on
$\caH_k$. Each reservoir is in thermal equilibrium at inverse
temperature $\beta_k$. One connects all heat reservoirs with a
small system $\sys$, modeled by a finite-dimensional Hilbert space
$\caH_\sys$ and a self-adjoint Hamiltonian $H_\sys$. The
connection between $\sys$ and reservoir $k$ is through a
self-adjoint coupling term $ H^{\mathrm{I}}_{\sys-k}$ on
$\caH_\sys \otimes \caH_k$. The coupling is controlled by a factor
$\la$ and the total Hamiltonian is \beq \label{def: formal ham}
 H_\la=H_\sys + \sum_{k }H_{k} + \la \sum_{k}
 H^{\mathrm{I}}_{\sys-k}
 \eeq

The dynamics of the full system is given by the unitary group
$U_t^{\la}=\e^{-\i t H_\la}$ on the Hilbert space $\caH
:=\caH_\sys \otimes \left[\otimes_{k } \caH_k \right]$. The
initial state $\rho_0$ is a product state on $\caH$ \beq
\label{def: initial state} \rho_0 = \mu_0 \otimes
\left[\mathop{\otimes}\limits_{k} \rho_{k,\beta_k} \right]\eeq
where the states $\rho_{k,\beta_k}$ are equilibrium states on
$\caH_k$ with respect to the Hamiltonians $H_k$ at inverse
temperatures $\be_k$ and $\mu_0$ is an arbitrary density matrix on
$\caH_\sys$.

In the thermodynamical limit, the states $\rho_{k,\beta_k}$ are no
longer density matrices and one needs the machinery of operator
algebras, see e.g.\ \cite{bratellirobinson,derzinski1}, to define
these states properly. We will completely ignore this problem
(since its solution belongs to standard knowledge in mathematical
physics), and we avoid writing ill-defined expressions by using
the notation \beq \rho_{k} [ A ] := \trace [\rho_{k} A] \qquad
\textrm{ with }A \textrm{ an operator and }  \rho_k  \textrm{ a
density matrix on } \caH_k \eeq The LHS can still be used when
$\rho_k$ is no longer a density matrix, but instead a state in the
sense of operator algebras\footnote{In fact, the situation is even
more complicated. For fermions, one can indeed find a subalgebra
of $\caB(\caH_k)$ -The Weyl-$\caC^*$-algebra- on which the state
$\rho_{k,\be_k}$ can be appropriately defined. For bosons, one has
to invoke  $W^*$-algebra's and the Araki-Woods representation to
give rigorous meaning to the model. } Note finally that models as
(\ref{def: formal ham}) have received a lot of attention lately in
mathematical physics. E.g.\ in \cite{jaksicogata} one proves the
Green-Kubo relations and Onsager reciprocity  in the spin-fermion
model at finite $\la
>0$.

\subsection{What do we mean by quantum fluctuations?}\label{sec: what do we
mean}

It is not a priori clear what one means by fluctuations of heat
currents, or fluctuations of the entropy production.  (See
\cite{allahverdyannieuwenhuizen} for an elaboration on that
question.)

One encounters at least three  approaches (see however
\cite{allahverdyannieuwenhuizen} for a different view), \ben
\item{Measure the energies $H_k$ in the beginning and at the end
of the experiments and make statistics of the difference of both
measurements. This approach yields the  FT
\cite{deroeckmaes,jarzynskiquantum,kurchanquantum,talknerlutzhanggi}.}
\item{Calculate the fluctuations of the operator $U_t H_k U_{-t}-
H_k$. In \cite{monnaitasaki}, a deviation from the fluctuation
theorem was established for  the fluctuations of a related
operator.} \item{In \cite{matsuitasaki}, one studies the
fluctuations of the \emph{relative modular operator} - a
$C$*-algebraic concept. These fluctuations satisfy the FT, but the
meaning of this relative modular operator is \emph{a priori}
unclear to us. } \een

The advantage of the first approach is that it is more relevant in
an experimental setup. A possible argument for the second would be
that it seems better suited for taking the thermodynamic limit,
since it avoids the measurements.  However, by slightly
reformulating the first approach, one can also meaningfully study
the thermodynamical limit (see also \cite{klich}), as we point out
now.

Assume for simplicity that the $H_{k}$ have discrete spectrum,
indicating that we have not taken the thermodynamic limit and let
$x \in X$ label a complete set of eigenvectors $| x
>$ of $(H_{k})_{k \in K}$ with eigenvalues $(H_{k})_{k
\in K}(x)$. Let the projectors $P_x$ stand for $1 \otimes
|x\rangle \langle x    |$ where $1$ stands for identity on
$\caH_\sys$. The probability of measuring an entropy production
$r$ is \beq \label{def: prob la} \bbP_{t,\la} (r)=
\mathop{\sum}\limits_{x,y \in X , \sum_{k}\beta_k
(H_k(y)-H_k(x))=r } \rho_0 \left[ P_x  U^\la_{-t} P_y   U^\la_{t}
P_x \right]
 \eeq
 The idea behind the formula is clear: measure (thereby projecting
on the eigenstates $x$), then switch on the time evolution
$U^{\la}_t$, finally measure again (projecting on the eigenstates
$y$). For later convenience, instead of studying the probability
distribution $\bbP_{t,\la}(r)$, we focus on the Laplace transform.
\beq F(\ka,t,\la,\rho_0):= \int_{\bbR} \d  \bbP_{t,\la}(r) \e^{ -
\ka r}
 \eeq
which, in the case when $\bbP_{t,\la}(r)$ is given by (\ref{def:
prob la}), reads \ \beq \label{alternative analogue}
F(\ka,t,\la,\rho_0):= \sum_{x,y \in X} \rho_0 \left[ P_x
U^\la_{-t} P_y   U^\la_{t} P_x \right] \, \e^{ - \ka \sum_{k}
\be_k \left( H_{k}(y) - H_{k}(x) \right) }
 \eeq
 We now use that the initial state $\rho_0$ is diagonal in
the basis $| x >$ to rewrite (\ref{alternative analogue}) into
\beq \label{alternative analogue2}F(\ka,t,\la,\rho_0)= \rho_0
\left[ U^\la_{-t} \e^{-\ka \sum_{k} \be_k H_{k}   } U^\la_{t}
\e^{\ka \sum_{k} \be_k H_{k}   } \right]
 \eeq
This last expression is perfectly suited for taking the
thermodynamic limit, since it  remains well-defined when the
operators $H_{k}$ have continuous spectrum and the state $\rho_0$
is not given by a density matrix. (Which doesnot mean that the
expression (\ref{alternative analogue2}) is necessarily finite.
After all, the physics could be such that the Laplace transform is
infinite for some $\ka$.)

\subsection{Derivation of the quantum fluctuation theorem}\label{sec: derivation of
ft}

If one defines the fluctuations through (\ref{alternative
analogue2}), FT is easily derived. Assume that the initial state
$\rho_0$ is given by \beq \rho_0= \frac{1}{d} \otimes \left[
\otimes_{k} \frac{ \e^{-\beta_k H_k}}{Z_k} \right], \qquad
Z_k:=\trace \e^{-\beta_k H_k}, \qquad d:= \dim \caH_\sys, \eeq
hence the initial state on the small system is the trace state, or
infinite temperature state. Assume further that the dynamics is
reversible, i.e.\ there is an anti-unitary time-reversal
involution $T$ which commutes with both the free and interacting
Hamiltonians \beq [T, H_k]=0 \qquad  [T, H_\la]=0 \qquad TT=1
\qquad T U_t^{\la}T=  U_{-t}^{\la} \eeq

Abbreviating $W= \sum_{k} \beta_k H_k$, using the above relations
and cyclicity of the trace, one calculates
\begin{eqnarray}
F(\ka,t,\la,\rho_0)=&   \trace \left[  \e^{-W} TT U^{\la}_{t} TT
\e^{- \ka W} TT U^{\la}_{-t} TT \e^{ \ka W} \right] (d \prod_k
Z_k)^{-1}& \label{der: ft1}
\\= &\trace \left[  \e^{-W} U^{\la}_{t} \e^{- (1-\ka) W} U^{\la}_{-t} \e^{(1-
\ka) W} \right] (d \prod_k Z_k)^{-1} & =   F(1-\ka,t,\la,\rho_0)
\label{der: ft2}
\end{eqnarray}
which is equivalent to (\ref{FT}) without the $o(t)$-correction
term.
 The above  derivation relies on a particular initial state for the
small system. It is known as a transient FT. Obviously, for large
$t$, one expects the initial state of the small system to be
irrelevant and (\ref{FT}) follows under some additional ergodicity
assumptions.

\subsection{Example: counting statistics for charge transport} \label{sec: ll}
As argued in Section \ref{sec: introduction}, it is interesting to
investigate to what extent the fluctuation symmetry is actually
present in known expressions for current statistics. We briefly
discuss the Lesovik-Levitov formula \cite{levitovjmp} for
electronic transport between leads. We base our treatment on a
neat version of the formula presented in \cite{klich} wherein the
electrons do not interact and the formula (\ref{ll}) can be
derived without any approximation. It is also assumed that all
Hilbert spaces are finite-dimensional (see \cite{avronbachmann}
for a discussion of the thermodynamical limit).

Let $\frh$ be the Hilbert space of one electron. The presence of
two leads is made explicit by splitting $\frh= \frh_1 \oplus
\frh_2$ where the spaces $\frh_1, \frh_2$ contain all the states
in respectively lead $1$, $2$. The projectors on $\frh_1 , \frh_2$
are denoted $1_{\frh_1},1_{\frh_2}$ and the decoupled one-particle
Hamiltonian is $h_1 \oplus h_2$. Let the inverse temperature be
$\beta$ and let the leads have respective chemical potentials
$\mu_1,\mu_2 $. Let $n$ be the occupation number operator
(dictated by Fermi-Dirac statistics) \beq n= \left( \e^{  \beta (
h_1 -\mu_1 1_{\frh_1})} \e^{\beta (h_2-\mu_2 1_{\frh_2}) } +1
\right)^{-1}
 \eeq The contact between the leads is modeled by a
scattering process with unitary scattering matrix $S$. We assume,
 that the scattering process conserves
total energy and that both scattering process and decoupled
evolution are reversible \beq \label{ass: counting} [S,h_1 \oplus
h_2]=0 \qquad Th_{1,2}T=h_{1,2} \qquad  T S T =S^* \eeq for some
anti-unitary operator $T$ modeling time-reversal.

At time $t=0$ and $t=\tau$, one measures the total charge in, say,
lead $1$ (For simplicity, we set the charge to unity, so that
`charge' just means  `number of particles'). Let $F(\ka)$ be the
laplace transform of the distribution of transported charge
$F(\ka)= \sum_{q =-\infty}^{+ \infty} p(q) \e^{-\ka q}$  where
$p(q)$ is the probability to find at time $t=\tau$  an excess
charge $q$, as compared to $t=0$.

The Lesovik-Levitov formula reads (see \cite{klich}, where the
formula is given for the Fourier transform, hence for $\ka \in \i
\bbR$):
 \beq  \label{ll} F(\ka)=  \textrm{det}
\left(  1+ n ( S^* \e^{- \ka 1_{\frh_1} } S \e^{\ka  1_{\frh_1}
}-1 ) \right) \eeq

The entropy production in this scenario is just $-
\be(\mu_1-\mu_2) q
 $ and hence one expects that FT implies the following symmetry
 (cfr.\ (\ref{der: ft1}-\ref{der: ft2}))
 \beq F(-\ka \be(\mu_1-\mu_2) ) =  F(-(1-\ka) \be(\mu_1-\mu_2)) \eeq
 That relation is easily verified by
 starting from (\ref{ll}),
¨
using (\ref{ass: counting}) and the fact that $ \textrm{det}(A B)=
 \textrm{det}(A)  \textrm{det}(B)$ for matrices
 $A,B$ (Recall that our one-fermion space is finite-dimensional).

\section{A good effective model: the master equation} \label{sec:
master}
\subsection{Construction of the master equation}
Having in mind the  Hamiltonian model introduced in Section
\ref{sec: qft}, we can construct a master equation that
approximates the evolution of a general density matrix $\mu_0$ on
the small system $\caH_\sys$. This master equation describes
essentially\footnote{That is, if we assume $H_\sys$ to be
nondegenerate. If it is degenerate, the picture is slightly more
complicated but the results still hold, see
\cite{deroeckmaesfluct}.} two effects: \ben \item{ Decoherence:
non-diagonal elements in the eigenbasis of $H_\sys$ vanish }\item{
The fluctuations of the diagonal elements, which evolve
independently of the off-diagonal elements.} \een In view of this
behavior, we choose to restrict attention to diagonal elements.
This allows for a  less technical presentation, as we can replace
the Lindblad master equation by a classical Markov process.

Let $e,e',\ldots$ be the eigenvalues of the system Hamiltonian
$H_\sys$ and let $P_e,P_{e'},\ldots$ be the projectors on the
corresponding eigenspaces which we assume one-dimensional.

Define transition probabilities for each seperate $k \in K$ as
\beq  \label{def: transition prob} p_k(e,e'):=  \int_{\bbR}\d t \,
\e^{\i t (e'-e)}    (1 \otimes \rho_{k,\be_k}) \left[ \e^{\i H_k
t}
 P_{e'}  H_{\sys-k}^{\mathrm{I}} P_{e} \e^{-\i H_k t} P_{e} H_{\sys-k}^{\mathrm{I}} P_{e'} \right]
   \eeq
where $1$ is the (unnormalized) density matrix $1$ on
 $\caH_\sys$.
The total  transition probabilities are hence given as \beq
\label{def: total trans prob} p(e,e')= \sum_{k } p_k(e,e') \eeq
The information about the temperatures $\beta_k$ is encoded into
the RHS of (\ref{def: transition prob}) through the state $\rho_k
$ and in fact one can show  \beq \label{boltzman
weights}\textrm{Reservoir $k$ is at thermal equilibrium at $\be_k$
} \Rightarrow
 p_k(e,e') / p_k(e',e) = \e^{\be_k (e-e') } \eeq

As said, the density matrix of the system is diagonal. We denote
the time-dependent diagonal elements by $\mu_e(t)$.  The (Pauli)
master equation now reads \beq  \label{eq: master}\frac{\d
\mu_e(t)}{\d t}= \sum_{e'}\left( \mu_{e'}(t) p(e',e)- \mu_e(t)
p(e,e')  \right) \eeq

\subsection{Derivation of the master equation}
The only information missing is in what sense this master equation
(\ref{eq: master}) approximates the true dynamics generated by
(\ref{def: formal ham}). Physically, one needs to perform the
Born-Markov approximation and the rotating wave approximation, on
which we will not comment here (see instead
\cite{alickiinvitation,breuerpetruccione}). Mathematically, one
can make the convergence to the master equation precise in the
so-called weak coupling limit: Let the initial state $ \rho_0 =
\mu_0 \otimes \left[\otimes_k \rho_{k,\beta_k} \right]$ be as in
(\ref{def: initial state}). Let $\mu^{\la}(t)$ be the density
matrix obtained by reducing the full state of the interacting
system at time $t$, i.e.\ \beq \mu^{\la}(t) = \trace_{\res}
\left[U_{t}^{\la}\, \mu_0 \otimes [\otimes_k \rho_{k,\beta_k}]\,
U_{-t}^{\la} \right] \eeq where $\trace_\res$ is a conditional
expectation. (If $\rho$ is a state on $\caH_\sys \otimes \caH_\res
$, then $\trace_\res \rho$ is a state on $\caH_\sys$ such that for
any operator $A$ on $\caH_\sys$: $\rho (A)= (\trace_R\rho) (A)$.
If $\rho$ is a density matrix, then $\trace_\res$ is just the
partial trace over the reservoirs.) The statement, rigorously
proven in \cite{davies1} under mild assumptions (about which we
will be more explicit in Section \ref{sec: result}), is

 \beq   P_e \mu^{\la}(\la^{-2}t) P_e \mathop{\longrightarrow}\limits_{\la \searrow 0} \mu_e(t)   \eeq

As is clear from the scaling and the expression (\ref{def:
transition prob}), the master equation emerges in second-order
perturbation theory. Hence we require that $\trace_\res [
H_{\sys-k}^{\mathrm{I}}]=0$ such that the first order term
perturbation term vanishes (which can always be achieved by
redefining $H_\sys$.)

\subsection{Guessing  path-wise entropy production} \label{sec:
guessing}

Given the very suggestive form of the transition probabilities
(\ref{def: transition prob}), one is inclined to supplement the
master equation in  the following way:

The paths ${\si}$ of the markov process defined by formula
(\ref{eq: master}) are sequences of energy levels $e_i$ with
transition times $t_i$ (jump from $e_{i-1}$ to $e_i$) with
${\bbP}_{e_0}^t$ the path probabilities up to time $t$
\begin{equation}\label{def: paths} {\si}=(e_0 ;t_1, e_1;t_2,e_2;
\ldots ; t_n,e_n) \qquad {\bbP}_{e_0}^t ({\sigma})= \e^{-\la_{e_0}
t_1}p(e_0,e_1) \e^{-\la_{e_1} (t_2-t_1)}
 \ldots p(e_{j-1},e_j)\e^{-\la_{e_{j}} (t-t_{j})}
\end{equation} where $j$ is the highest index such that $t_j <t$ and
$\la_{e}= \sum_{e'}p(e,e')$ are the escape rates.  The subscript
$e_0$ in ${\bbP}_{e_0}^t $ indicates the the process has been
started from $e_0$.

Clearly, to speak about currents and entropy production, we need
more information; we need to know which bath $k$ \emph{triggered}
the transition. Hence, we need a process with paths
$\tilde{\sigma}$ and probabilities $\tilde{ \bbP}^t_{e_0}
(\tilde{\sigma})$; \beq \label{def: extended paths}\tilde{\sigma}=
(e_0 ;t_1,k_1, e_1;t_2,k_2,e_2; \ldots ; t_n,k_n,e_n) \qquad
\tilde{ \bbP}^t_{e_0} (\tilde{\sigma})= \e^{-\la_{e_0}
t_1}p_{k_1}(e_0,e_1) \e^{-\la_{e_1} (t_2-t_1)} \ldots
p_{k_j}(e_{j-1},e_j)\e^{-\la_{e_j} (t-t_j)}\eeq We simply define a
new process corresponding to the pathspace measure (\ref{def:
extended paths}). Remark that obviously, \beq \sum_{\tilde{\si}
\rightarrow \si  } \tilde{\bbP}^t_{e_0}(\tilde{\si}) =
{\bbP}^t_{e_0} (\sigma) \eeq where the sum is over all
$\tilde{\si}$ which reduce to $\si$ by omitting all $k_i$.

One can just as well calculate the matrix elements ${\mu}_e(t)$
from the path probabilites $\tilde{\bbP}$ as from ${\bbP}$: \beq
{\mu}_e(t)= \sum_{\tilde{\si}, e_{n(\tilde{\si})}=e}\tilde{
\bbP}^t_{\mu(0)}(\tilde{\si})=  \sum_{\si, e_{n(\si)}=e}
\bbP^t_{\mu(0)}(\si)  \eeq where \beq
\tilde{\bbP}_{\mu(0)}^t=\sum_{e_0} \mu_{e_0}(0)
\tilde{\bbP}_{e_0}^t, \qquad \bbP_{\mu(0)}^t=\sum_{e_0}
\mu_{e_0}(0) \bbP_{e_0}^t  \eeq and with $n(\tilde{\si}), n(\si) $
being the number of jumps in the paths $\tilde{\si},\si$.
 Hence, we have
extended our pathspace probability measure $\bbP^t$ into a
pathspace probability measure $ \tilde{\bbP}^t$. In the theory of
open quantum systems, one calls such an extension an
\emph{unraveling} and the paths $\tilde{\si}$ are called
\emph{quantum trajectories}, see
\cite{carmichael,breuerpetruccione} for an overview and
\cite{choughcarmichael} for  similar reasoning.

Having the pathspace probability measure $\tilde{\bbP}^t$ at our
disposal, we can do all the manipulations which are familiar from
classical markov processes.  The function on paths $\tilde{\si}$
\beq \label{def: ent prod}w^t(\tilde{\si})= -\sum_{i=1}^n
\be_{k_i}(e_{i}-e_{i-1})
  \eeq
which we identify as the path dependent entropy production, is
equal to the ratio of probabilities of forward and backward paths;
\beq \label{radon} w^t (\tilde{\si})  = \log \frac{\d
\tilde{\bbP}_{\mu_0} (\tilde{\si}) }{\d \tilde{\bbP}_{\mu_t}
(\theta \tilde{\si}) } \quad \textrm{ with }
 \theta \tilde{\si}  =  (e_n ;t-t_n,k_n,e_{n-1} ; \ldots
;t-t_2,k_2,e_2 ; t-t_1,k_1,e_0)
 \eeq
as follows from (\ref{boltzman weights}). From (\ref{radon}), we
retrieve the FT (\ref{FT}), or its Laplace-transformed version
\beq \label{statement FT} \log\int \d
\tilde{\bbP}^t_{\mu_{0}}(\tilde{\si}) \e^{-\ka w^t(\tilde{\si})} =
\log \int \d \tilde{ \bbP}^t_{\mu_{0}}(\tilde{\si}) \e^{-(1-\ka)
w^t(\tilde{\si})} +o(t) \eeq Obviously, one can push on and prove
the Green-Kubo relations, Onsager reciprocity and strict
positivity of the entropy production from (\ref{statement FT}).
This is discussed at length in \cite{deroeckmaesfluct}.


At this point, one should  note that this \emph{unraveling} and
the construction of the pathspace measure $\tilde{\bbP}^t$ was a
product of our intuition.  Hence, it not a priori clear whether
the fluctuations of entropy production which we can calculate
through formula (\ref{def: ent prod}) and the measure
$\tilde{\bbP}^t$ coincide with the entropy fluctuations in the
original model. This problem motivates the following section.

\begin{remark}\label{rem: unravel}

However intuitive the above reasoning, one should keep in mind
that we have used more information than is contained in the bare
master equation (\ref{eq: master}). In particular, we used
(\ref{def: total trans prob}). To put things even sharper: given a
master equation without any additional information, one could
associate to it baths in different ways, thereby obtaining
completely different expressions for the entropy production.
\end{remark}

\section{Linking the microscopic fluctuations with the quantum trajectory fluctuations }\label{sec: result}

The idea of this section is to justify the guesses in Section
\ref{sec: guessing}. We do this by connecting fluctuations of the
entropy production in the Hamiltonian model, as in expression
(\ref{alternative analogue2}) to   the fluctuations calculated by
the probability measure $\bbP_{\mu_0}^t$. This connection is made
sharp in the regime in which the master equation is derived, i.e.\
in the weak coupling limit.   To be as concrete as possible, we
make explicit the model which was outlined in the beginning of
Section \ref{sec: qft}.

 Let for each $k$, $\caH_k$ be a fermionic\footnote{One
could just as well take bosons here, but then formulas like
(\ref{def: ham fluct}) need more elaboration to get rigorous
meaning, see also the footnote in the introduction to Section
\ref{sec: qft}.} Fock space, i.e.\ \beq \caH_k := \bbC \oplus
L^2(\bbR^d) \oplus L^2(\bbR^d) \otimes_a L^2(\bbR^d) \oplus \ldots
\eeq where the $\otimes_a$ denotes the anti-symmetrized tensor
product, which accounts for the fermion statistics. Let the
Hamiltonian $H_k$ describe free fermions with one-particle
dispersion relations $h_k(q)=h_k(|q|)$ where $q\in \bbR^d$ is the
momentum, i.e.\ \beq H_k = \int_{\bbR^d} \d q \, h_k(|q|) a^*_k
(q) a_k (q) \eeq with $a_k (q), a^*_k (q) $ creators/annihilators
of fermions in reservoir $k$ with momentum $q$, that is \beq
\{a_k(q),a^*_{k'}(q') \}=  \de_{k,k'} \de (q-q') \eeq

 Let now the
interaction $H^{\mathrm{I}}_{\sys-k}$ be given as \beq
H^{\mathrm{I}}_{\sys-k}= \int_{\bbR^d} \d q \, f_k(q) D_k \otimes
a^*_k (q) + \bar{f}_k(q)D^*_k \otimes a_k (q) \eeq for some
operators $D_k$ on $\caH_\sys$ and coupling functions $f_k \in
L^2(\bbR^d)$. The thermal states $\rho_{k,\beta_k}$ on $\caH_k$
are defined as quasi-free states determined by their two-point
correlation function (determined by the Fermi-Dirac distribution)
\beq \rho_{k,\beta_k} [a_k^*(g') a_k(g)] = \langle g, (1+
\e^{\be_kh_k })^{-1} g' \rangle \eeq

We recall the definition of $F(\ka,t,\la,\rho_0)$ in
(\ref{alternative analogue2}) as the Laplace transform of the
distribution of entropy production; \beq \label{def: ham fluct}
F(\ka,t,\la,\rho_0)= \rho_0 \left[ U^\la_{-t} \e^{- \ka  \sum_{k}
\be_k H_{k}   } U^\la_{t} \e^{ \ka \sum_{k} \be_k H_{k}   }
\right]
 \eeq
 where again, as in (\ref{def: initial state}), $\rho_0=
\mu_0 \otimes \left[ \otimes_{k} \rho_{k,\beta_k} \right] $ with
$\mu_0$ an arbitrary density matrix on $\caH_\sys$.

We can now state the main message of this paper.

 \begin{tabular}{ll} \qquad & \qquad   \\
  \qquad \qquad &
 \hbox{\fbox{\begin{minipage}[c]{0.80\textwidth}
\emph{ The generating function $F(\ka,t,\la,\rho_0)$ (\ref{def:
ham fluct}) from the microscopic Hamiltonian model converges in
the weak coupling limit to the generating function defined by
unravelings, \[ F(\ka,t,\mu_0) = \int_{\bbR} \d r \e^{-\ka r}
\tilde{\bbP}_{\mu_0}^t \left[ w^t = r \right]
 \]
where $\tilde{\bbP}_{\mu_0}^t$ is the measure constructed in
Section \ref{sec: guessing}. That is, under some conditions (see
below) and for all  $t
>0$ and initial density matrix $\mu_0$ on $\caH_\sys$,
\[  F(\ka,\la^{-2}t,\la,\rho_0) \mathop{\longrightarrow}\limits_{\la
\searrow 0} F(\ka,t,\mu_0)
\]
}
  \end{minipage} }} \\
  \qquad & \qquad

\end{tabular}

Rephrasing the fluctuation relation within the Hamiltonian model
and taking $t \uparrow \infty$ one gets :
 \beq \lim_{t \nearrow +\infty} \lim_{\la \searrow 0}
t^{-1}\log{F(\ka,\la^{-2}t,\la,\rho_0)} = \lim_{t \nearrow
+\infty} \lim_{\la \searrow 0}
t^{-1}\log{F(1-\ka,\la^{-2}t,\la,\rho_0)}
 \eeq
which is a mathematical statement about large deviation generating
functions.

The limit $\la \searrow 0$ can be omitted altogether from the
above statement since the FT holds without small-coupling
approximation (see \cite{deroeck} for a proof of existence of the
limit $t \nearrow +\infty$ in that case).

These statements are essentially  a rephrasing of technical
results in \cite{derezinskideroeck2}, and they are more
extensively commented upon in \cite{derezinskideroeckmaes}. The
results are proven by extending the convergence in the weak
coupling limit to the full system (including reservoirs). One
proves that the full Hamiltonian evolution converges to a quantum
stochastic evolution. Similar results (though in a weaker form,
not permitting the conclusions of this paper) have been proven in
\cite{duemcke} and \cite{accardifrigerio}.

To make the discussion complete, we give an example of a set of
assumptions that is sufficient to prove the above statements.

Define the set of Bohr frequencies $\caF:= \{ e-e' \, |  e,e' \in
\sp H_\sys \}$ and assume that \ben
\item{The dispersion functiosn $h_k: \bbR^+ \mapsto \bbR^+$ are monontone increasing,
and continuously differentiable with nonzero  derivative in
$h_k^{-1}(\caF)$. The coupling functions $f_k$ are also continuous
in $h_k^{-1}(\caF)$. Write $h_k$ for the self-adjoint operator
associated to
  the dispersion function $h_k$ by $(h_k g)(q)=h_k(|q|)g(q)$ for
  $g \in L^2(\bbR^d)$.
  }

\item{ For
 $\ga=0,\ka$, the functions \beq  \bbR \ni t
 \to \left\{\begin{array}{l}
\rho_{k,\be_k} [a(\e^{-\i t h_k}\e^{-\ga \be_k  h_k} f) a^*( f)] \\
\rho_{k,\be_k} [a^*(\e^{-\i t h_k}\e^{\ga \be_k h_k} f) a(f)]
\end{array}\right. \eeq are bounded and $L^1$-integrable
   }
 \een

\section{Conclusion}

We investigated the question of generalization of FT to quantum
systems. We remarked in Section \ref{sec: derivation of ft} that
in a microscopic (=Hamiltonian) framework, one can unambiguously
state FT.  In practice, it is more common to start from a
mesoscopic (=effective, nonhamiltonian) description of an open
quantum system.   In Section \ref{sec: guessing}, we discussed a
very straightforward and successful approach  for the dynamics
given by a specific master equation; one can guess what are
fluctuations of entropy production and one can state FT.  Finally,
in Section \ref{sec: result}, we state the link between
fluctuations of the entropy production, defined in the microscopic
model in Section \ref{sec: qft} and the fluctuations which are the
result of the guess-work in Section \ref{sec: master}, which hence
turns out to be good guess-work.

\bibliographystyle{plain}
\bibliography{procbrussel3}

\end{document}